\documentclass[aip,apl,amsmath,amssymb,reprint,floatfix]{revtex4-1}
\usepackage{graphicx,amsmath}
\usepackage{hyperref}
\usepackage{times}

\newcommand{\rnd}[1]{\left(  #1 \right)}

\newcommand{\unit}[1]{\,\mathrm{#1}}
\newcommand{\comment}[1]{}

\begin{document}

\title{A chemical turnstile}
\thanks{Copyright (2005) American
  Institute of Physics. This article may be downloaded for personal
  use only. Any other use requires prior permission of the author and
  the American Institute of Physics. The following article appeared in
  \textit{Appl.\ Phys.\ Lett.\ }\textbf{87}:143110 (2005) and may be
  found at \url{http://link.aip.org/link/?apl/87/143110}}

\author{W. T. Lee} \affiliation{Mineralogisch-Petrographisches
  Institut, Universit\"at Hamburg, Grindelallee 48, D-20146 Hamburg,
  Germany.}  \affiliation{Department of Earth Sciences, University of
  Cambridge, Downing Street, Cambridge, CB2 3EQ, UK.}

\author{E. K. H. Salje} \affiliation{Department of Earth Sciences,
  University of Cambridge, Downing Street, Cambridge, CB2 3EQ, UK.}

\begin{abstract}
A chemical turnstile is a device for transporting small, well-characterised doses of atoms from one location to another. A working turnstile has yet to be built, despite the numerous technological applications available for such a device. The key difficulty in manufacturing a chemical turnstile is finding a medium which will trap and transport atoms. Here we propose that ferroelastic twin walls are suitable for this role. Previous work shows that twin walls can act as two-dimensional trapping planes within which atomic transport is fast. We report simulations showing that a stress-induced reorientation of a twin wall can occur. This behaviour is ideal for chemical turnstile applications.

\end{abstract}

\pacs{66.30.Pa, 62.20.Dc, 81.07.--b}

\maketitle

\begin{figure}
\begin{center}
\includegraphics[]{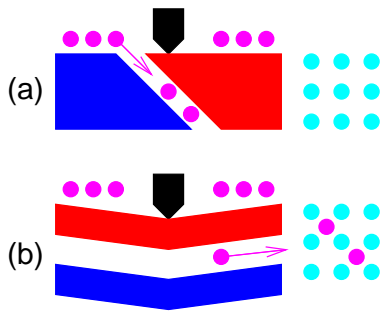}
\end{center}
\caption{Schematic of a chemical turnstile. In (a) atoms diffuse from the
  source into the trapping plane. In (b) the trapping plane is reoriented and
  the atoms allowed to diffuse out into the host material.}
\label{TurnstileCartoon}
\end{figure}

There are many areas of technology in which the ability to deliver a
chemical dose in small, well characterised quanta would be
invaluable. Some examples can be found in biomedical technology, for
the controlled release of a drug, as well as in nanotechnology for
precision doping. For these applications we propose the idea of a
chemical turnstile. The electronic equivalent, the electron turnstile,
is already well known.\cite{Geerligs1990} A schematic of a chemical
turnstile is shown in Fig.~\ref{TurnstileCartoon}. It possesses a
trapping plane which captures diffusing ions. Diffusion within the
plane occurs easily, but diffusion from within the plane into the bulk
crystal is strongly inhibited. In the first stage of the turnstile
operation (Fig.~\ref{TurnstileCartoon}a) this trapping plane connects
the upper and lower surfaces of the crystal. All the available sites
within the plane will be occupied by dopant atoms: the number of such
sites will define the dose quantum of the device. In the second stage
of operation (Fig.~\ref{TurnstileCartoon}b) the trapping plane is
reoriented so that it connects the lateral surfaces. This allows the
dopant atoms to diffuse out into the host material. To build up the
required dose, the turnstile can be repeatedly cycled through these
two stages of operation.

A working chemical turnstile has yet to be built: the major obstacle
is finding a material with a trapping plane which has the properties
required. These are
\begin{itemize}
\item{Two dimensional confinement to a small number of trapping sites.}
\item{Fast exchange between trapping sites}
\item{Controllable reorientation of the trapping plane}
\end{itemize}
Previous experiments have shown that ferroelastic crystals posses the
first two properties. Ferroelastic crystals exhibit anisotropic
strains as a result of phase transitions---their microstructures show
numerous twin walls between different variants of the low temperature
phase.\cite{Salje1990} These twin walls have a number of properties
which make them very attractive as components in nanoscopic
devices.\cite{Salje1999} Evidence of confinement within twin walls
comes from direct measurements, simulations and from the observation
of experimental effects which can be attributed to the trapping of
atomic species by the wall. Examples of direct measurements showing
confinement within twin walls include sodium diffusion experiments in
tungsten oxide and ion microprobe measurements of anorthoclase twin
walls.\cite{Aird2000,Camara2000} These measurements also showed that
diffusion within the twin wall was significantly faster than bulk
diffusion. Simulations showing chemical confinement within twin walls
include simulations of generic models, sodium ions in quartz domain
walls and trapping of oxygen vacancies within twin walls of calcium
titanate and lead
titanate.\cite{Lee2002a,Calleja2001,Calleja2003,He2003} Indirect
experimental evidence for the confinement of atoms within twin walls
is given by, for example, the twin memory effect.\cite{Frondel1945,
  Voronkova1993} The small width of twin walls, typically a few unit
cells at most,\cite{Hayward1996,Shilo2004} shows that the number of
trapping sites within the twin wall is small compared with the total
number of sites in the crystal. Thus there is ample evidence that
ferroelastic crystals satisfy the first two requirements. The third
requirement is more difficult to fulfil. Experimental evidence from
combined \textit{in situ} X-ray diffraction and dynamical mechanical
analysis shows that domain walls in lanthanum aluminate rotate in
response to an applied stress.\cite{Harrison2004} This, in itself, is
not a sufficiently large orientational change to be useful in a
chemical turnstile. Here we report simulations of the response of a
twin wall to an applied bending stress which show a complete
reorientation of the twin wall under a sufficiently large bending
stress. This change is large enough for chemical turnstile
applications.

\begin{figure}
\begin{center}
\includegraphics{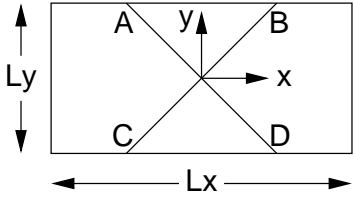}
\end{center}
\caption{Geometry used in equations~\ref{FreeEnergy} and~\ref{Stress}. The
  lines AD and CB show the permissible twin wall orientations in the bulk
  material.}
\label{geometry}
\end{figure}

To simulate the domain wall we used the method of finite elements, using the
following free energy to describe the system
\begin{equation}
\label{FreeEnergy}
F=\dfrac{a}{2Q_0}\rnd{\epsilon_2-Q_0^2}^2 +\dfrac{c_1}{2}\epsilon_1^2
+\dfrac{c_2}{2}\epsilon_{xy}^2 +\dfrac{g}{2}\rnd{\nabla \epsilon_2}^2
\end{equation}
where $\epsilon_1=\epsilon_{xx}+\epsilon_{yy}$ and
$\epsilon_2=\epsilon_{xx}-\epsilon_{yy}$. The geometry of the system is shown
in Fig.~\ref{geometry}. The ground state consists of a single domain with
$\epsilon_1=0$, $\epsilon_{2}=\pm Q_0$ and $\epsilon_{xy}=0$. Twin walls in
the system should form $45^{\circ}$ angles to the $x$ and $y$ axes. We will
describe the details of the implementation of our model
elsewhere.\cite{Lee2004a} To this system we apply a bending stress of
the form
\begin{equation}
\label{Stress}
\sigma_{xx}=\sigma_0 \dfrac{2y}{L_y} 
\end{equation}

\begin{figure}
\begin{center}
\includegraphics{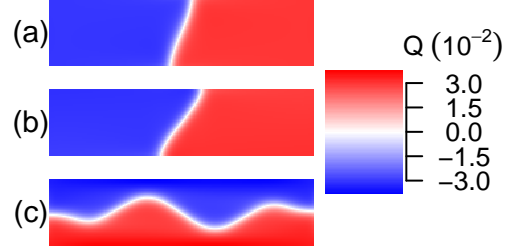}
\end{center}
\caption{Wall trajectories under different bending stresses. In (a) there is
  no bending stress, in (b) a small bending stress has been applied, and in
  (c) the bending stress is larger than the critical value.}
\label{trajectory}
\end{figure}

Our simulations show that for small values of the bending stress there
is a rotation of the twin wall, in agreement with experiments quoted
in ref.~\onlinecite{Harrison2004}. When the stress reaches a critical
value, one twin is destabilised at the upper and lower surfaces. The
twin wall is forced to disconnect from these surfaces and to follow a
path confined to the low stress region in the centre of the crystal,
as shown in Fig.~\ref{trajectory}. In part~(a) the wall is shown in
its stress free configuration. The wall is rotated away from the ideal
$45^{\circ}$ orientation due to the finite size of the
simulation. In~(b) a bending stress small compared to the critical
bending stress has been applied, and the domain wall has rotated. In
part~(c) a stress larger than the critical bending stress has been
imposed and the twin wall is no longer stable on the upper and lower
surfaces. These results show that ferroelastic twin walls also fulfil
the final requirement needed to work in a chemical turnstile.

For chemical turnstile applications, one important question is, how
much material will be transported in a single cycle of the turnstile?
i.e.\ what is the dose quantum? Consider a crystal of cross-section
$2\, \mu \mathrm{m}$ by $0.2\, \mu \mathrm{m}$. Take the unit cell
size to be of the order of $0.5\unit{nm}$ and assume one dopant site
per unit cell. Finally, assume that the twin wall width is
approximately one unit cell. This gives a dose quantum of the order of
$10^5$ atoms. A device of nanoscopic size would reduce the dose
quantum even further to give a truly atomic resolution.

It seems likely that, in the near future, ferroelastic twin walls will
be used in a number of nanotechnological applications.\cite{Salje1999}
These will use both the static properties of the twin walls
e.g.\ exploiting the confined superconductivity of WO$_3$
walls~\cite{Aird1998} and their dynamic properties, such as the
reorientation under an applied bending stress described here. The
chemical turnstile is an important example of the latter type of
application.


\begin{thebibliography}{16}
\bibitem{Geerligs1990} L.\ J.\ Geerligs, V.\ F.\ Anderegg,
  P.\ A.\ M.\ Holweg, J.\ E.\ Mooij, H.\ Pothier, D.\ Esteve,
  C.\ Urbina, and M.\ H.\ Devoret, Phys.\ Rev.\ Lett.\ \textbf{64},
  2691 (1990).
\bibitem{Salje1990} E.\ K.\ H.\ Salje, \textit{Phase Transitions in
  Ferroelastic and Co-Elastic Crystals} (Cambridge University Press,
  Cambridge, UK, 1990).
\bibitem{Salje1999} E.\ K.\ H.\ Salje, A.\ Aird, K.\ R.\ Locherer,
  S.\ A.\ Hayward, J.\ Novak, and J.\ Chrosch, Ferroelectrics
  \textbf{223}, 1 (1999).
\bibitem{Aird2000} A.\ Aird and E.\ K.\ H.\ Salje, Eur.\ Phys.\ J.\ B
  \textbf{15}, 205 (2000).
\bibitem{Camara2000} F.\ C\'amara, J.\ C.\ Doukhan, and
  E.\ K.\ H.\ Salje, Phase Transitions \textbf{71}, 227 (2000).
\bibitem{Lee2002a} W.\ T.\ Lee, E.\ K.\ H.\ Salje, and U.\ Bismayer,
  Phase Transitions \textbf{76}, 81 (2003)
\bibitem{Calleja2001} M.\ Calleja, M.\ T.\ Dove, and
  E.\ K.\ H.\ Salje, J.\ Phys.: Condens.\ Matter \textbf{13}, 9445
  (2001)
\bibitem{Calleja2003} M.\ Calleja, M.\ T.\ Dove, and
  E.\ K.\ H.\ Salje, J.\ Phys.: Condens.\ Matter \textbf{15}, 2301
  (2003).
\bibitem{He2003} L.\ He and D.\ Vanderbilt, Phys.\ Rev.\ B
  \textbf{68}, 134103 (2003).
\bibitem{Frondel1945} C.\ Frondel, Am.\ Mineral.\ \textbf{30}, 447
  (1945).
\bibitem{Voronkova1993} V.\ I.\ Voronkova and T.\ Wolf, Physica C
  \textbf{218}, 175 (1993).
\bibitem{Hayward1996} S.\ A.\ Hayward, J.\ Chrosch, E.\ K.\ H.\ Salje,
  and M.\ A.\ Carpenter, Eur.\ J.\ Mineral.\ \textbf{8}, 1301 (1996).
\bibitem{Shilo2004} D.\ Shilo, G.\ Ravichandran, and K.\ Bhattacharya,
  Nat.\ Mater.\ \textbf{3}, 453 (2004).
\bibitem{Harrison2004} R.\ J.\ Harrison, S.\ A.\ T.\ Redfern,
  A.\ Buckley, and E.\ K.\ H.\ Salje, J.\ Appl.\ Phys.\ \textbf{95},
  1706 (2004).
\bibitem{Lee2004a} Unpublished.
\bibitem{Aird1998} A.\ Aird and E.\ K.\ H.\ Salje, J.\ Phys.:
  Condens.\ Matter \textbf{10}, L377 (1998).
\end{thebibliography}

\end{document}